\documentstyle[astrobib,epsfig]{mn}

\begin{document}
\def\bi#1{\hbox{\boldmath{$#1$}}}

\newcommand{\beq}{\begin{equation}}
\newcommand{\eeq}{\end{equation}}
\newcommand{\beqa}{\begin{eqnarray}}
\newcommand{\eeqa}{\end{eqnarray}}

\newcommand{\lexp}{\mathop{\langle}}
\newcommand{\rexp}{\mathop{\rangle}}
\newcommand{\rexpc}{\mathop{\rangle_c}}

\newcommand{\cmm}{$ \xi_{mm} $}
\newcommand{\cgg}{$ \xi_{gg} $}
\newcommand{\cgm}{$ \xi_{gm} $}
\newcommand{\dd}{\delta}
\newcommand{\g}{\gamma_t}
\newcommand{\gam}{\gamma}
\newcommand{\kk}{\kappa}
\newcommand{\tht}{\theta}
\newcommand{\tl}{\tilde}
\newcommand{\bm}{\boldmath}
\newcommand{\vecr}{\vec{r} }
\newcommand{\DD}{\Delta^2}
\newcommand{\BB}{{\rm M}_{B}}
\newcommand{\bb}{{\rm m}_{B}}
\newcommand{\be}{\begin{equation}}
\newcommand{\ee}{\end{equation}}
\newcommand{\CCF}{cross-correlation function}
\newcommand{\kunit}{\; h {\rm Mpc}^{-1}}
\newcommand{\runit}{\; h^{-1} {\rm Mpc}}
\newcommand{\path}{./}

\title{Virial masses of galactic halos from 
galaxy-galaxy lensing: theoretical modeling and application to SDSS}

\author[Jacek Guzik \& Uro\v s Seljak ]
{Jacek Guzik$^{1,2}$ \& Uro\v s Seljak$^1$\\
$^1$Department of Physics, Jadwin Hall, Princeton University, Princeton, NJ 08544 \\
$^2$Astronomical Observatory, Jagiellonian University,
	 Orla 171, 30-244 Krak\'ow, Poland 
}

\pubyear{2001}

\maketitle

\begin{abstract}
We present a theoretical analysis of galaxy-galaxy lensing in the 
context of halo models with CDM motivated dark matter profiles. The
model enables us to separate between the central galactic and noncentral 
group/cluster contributions. We apply the model to the recent SDSS measurements
with known redshifts and luminosities of the lenses. This allows one to  
accurately model the mass distribution of a local galaxy population around and 
above $L_{\star}$.  We find that virial mass of $L_{\star}$ 
galaxy is $M_{200}=(5-10)\times 10^{11}h^{-1}M_{\sun}$ depending on 
the color of the galaxy. This value varies significantly with galaxy 
morphology with $M_{\star}$ for late types being a factor of 10 lower in $u'$, 
7 in $g'$ and a factor of 2.5-3 lower in $r'$, $i'$ and $z'$ relative to 
early types. Fraction of noncentral galaxies in groups and clusters is 
estimated to be below 10\% for late types and around 30\% for early types. 
Using the luminosity dependence of the signal we find that for early types the
virial halo mass $M$ scales with luminosity as $M \propto L^{1.4 \pm 0.2}$ in 
red bands above  $L_{\star}$. This shows that the virial mass to light ratio 
is increasing with luminosity for galaxies above $L_{\star}$, as predicted by 
theoretical models. The virial mass to light ratio in $i'$ band is
$17(45)hM_{\sun}/L_{\sun}$ at $L_{\star}$ for late (early) types.
Combining this result with cosmological baryon fraction one finds that
70(25)\%$h^{-1}\Upsilon_i\Omega_m/12\Omega_b$ of baryons within $r_{200}$ 
are converted to stars at $L_{\star}$, where $\Upsilon_i$ is the stellar mass 
to light ratio in $i'$ band. This indicates that both for early and late 
type galaxies around $L_{\star}$ a significant fraction of all the baryons 
in the halo is transformed into stars.
\end{abstract}



\section{Introduction}

Weak lensing by matter along the line of sight between the
source and the observer shears the images
of the background galaxies, 
inducing ellipticity distortions (see \citeNP{2001PhR...340..291B} 
for a review of weak lensing).
Although away from rich clusters the effect is too small to be
detectable for individual galaxy lenses, it can be measured 
statistically as a
function of relative separation from the galaxy. 
This requires averaging over the tangential ellipticities of all the background 
galaxies relative to the lens and over all the 
lenses (galaxies which are in the foreground). Until recently 
this averaging, named galaxy-galaxy (g-g) lensing, was done as a function of 
apparent angular position in the sky (\shortciteNP{1984ApJ...281L..59T}, 
\citeNP{1996ApJ...466..623B}, \shortciteNP{1998ApJ...503..531H}, 
\shortciteNP{2000AJ....120.1198F}, 
\citeNP{2001ApJ...551..643S}), 
so a signal at a given 
angular separation could be either coming from a small radial distance
of a nearby lens or from a large distance of a far lens. This 
made the theoretical interpretation of the data rather involved.  
First attempt to use distances was by \shortciteN{2001ApJ...555..572W}, 
which however only had limited photometric information and so could only 
obtain reliable distances to early type galaxies. 

Recent study of galaxy-galaxy lensing by 
SDSS collaboration \shortcite{2001astro.ph..8013M} is a significant step forward in 
the study of galaxy-galaxy lensing. 
The spectroscopic 
sample of more than 35,000 lensing galaxies and 3.6 million background galaxies  
is large enough to allow one a detailed study 
of the relation between mass and light for several luminosity bands and 
morphological types. 
Since the distances for lens galaxies are known 
one can study the strength of the signal as a function 
of proper radial separation from the lens. 
In addition, because the survey is 
shallow, the redshift distribution of background
galaxies is known from the deeper spectroscopic surveys (e.g. 
\citeNP{1995ApJ...455...50L}).
In combination with above this means that the mean critical density is known
for every lens, so one can average over the proper projected mass density rather
than the shear itself. 
This fact
greatly simplifies the theoretical analysis, since one can now 
measure the actual projected density as a function of galaxy luminosity and 
proper radial transverse distance from the galaxy.
In addition, since the lens sample is at low redshift
(mean $\bar{z}\sim 0.1$) redshift evolution is small and k-corrections 
are relatively reliable in red bands, further simplifying 
the theoretical interpretation. 

In this paper we want to connect the SDSS observational results to the
theoretical models in the context of our current understanding of
galaxy formation models within the CDM paradigm. 
We will model the dark matter halos with CDM type of halo profiles 
(NFW profile; \citeNP{1997ApJ...490..493N}), 
where the slope is gradually changing from the inner slope between -1 to 
-1.5 to the outer slope of -3. Although
other profiles have been proposed that differ significantly from NFW
in the inner parts of the halo, they agree well with NFW in the
outer parts \shortcite{2001ApJ...554..903K}. NFW profile should be contrasted 
to the truncated singular 
isothermal sphere with a constant slope -2 out to a fixed radius
which was often used used in the past work on g-g lensing. 
Our main goal is to determine the virial mass
of the halo and its relation to the 
luminosity of the galaxy. This is important for theoretical models of 
galaxy formation, since it 
is usually assumed that only baryons within the virial radius are able to 
condense and form stars. By determining the virial mass for a given galaxy 
luminosity one can thus directly determine the efficiency 
of star formation in typical galactic halos. 
Luminosity dependence of 
the galaxy-galaxy lensing signal allows one to determine this as a function 
of halo mass, while morphology subsamples can determine it as 
a function of morphological type. 
By comparing the radial dependence of the signal with the theoretical 
models one can also determine the fraction of galaxies that live in 
larger halos such as groups and clusters. This is another parameter that 
can distinguish between the different galaxy formation models. Finally, 
comparison of high density sample to the field sample allows one 
to study the effects of dense enviroments, such as tidal 
stripping, on the dark matter profile and mass to light ratios. 
Thus g-g lensing allows one to make detailed tests 
of the galaxy formation models 
(e.g. \shortciteNP{1999MNRAS.303..188K},
\shortciteNP{2000MNRAS.311..793B}, \citeNP{1999MNRAS.310.1087S}). 
In principle g-g lensing can also 
allow a direct determination of the dark matter halo profile, although 
as we will show here at present the data do not have enough power
to strongly constrain this.

An important issue in the theoretical analysis of g-g lensing
is how to separate the contribution from the individual halo of the 
galaxy  
from that of the neighbouring galaxies or larger mass concentrations such as 
groups and clusters. The former should dominate on small scales while
the latter on large scales. 
Previous work used the additional information obtained 
from the galaxy clustering to remove this contribution, but in general 
such analysis relies on the assumption that all the mass is associated 
with galaxies, which is invalid on scales where groups and clusters 
become dominant. 
Most of the mass
in these systems is in a diffuse form and only about 5-10\% is expected to 
be attached to 
individual galaxies (e.g. \shortciteNP{2001MNRAS.328..726S}, 
\shortciteNP{2000ApJ...544..616G}, \shortciteNP{2001MNRAS.321..559B}). 
This means that these systems cannot be 
modeled using the galaxies as the mass tracers, which only accounts for 
a small fraction of the total mass. In fact, theoretical models presented 
in this paper suggest
the contribution from these group and cluster halos 
dominates the signal on scales above 200$h^{-1}$kpc (see also
\citeNP{2000MNRAS.318..203S})
and has to be carefully modeled to account for it. 

Given that corrections to the profile on large scales are difficult to 
determine from the data directly one has to turn to theory for
guidance.
Realistic theoretical modeling of g-g lensing must combine galaxy formation models 
and dark matter models. One way is to use semi-analytic 
or hydrodynamic models
of galaxy formation and combine them with N-body simulations
(\citeNP{2001MNRAS.321..439G}; \citeNP{2001astro.ph..7023W}). 
This has the advantage of having a realistic distribution 
of dark matter and galaxies, but suffers from the lack of 
force and mass resolution. In addition, this approach by 
itself does not allow for a fast 
exploration of parameter space. The alternative approach is to  
use the recently popularized halo model applied to galaxies 
(\citeNP{2000MNRAS.318..203S}, 
\citeNP{2000MNRAS.318.1144P},
\shortciteNP{2001ApJ...546...20S}, \citeNP{2001astro.ph..9001B}), 
which takes into 
account both the individual halo profiles (for galaxies either at 
the centers of the galactic halos or distributed within 
larger groups and clusters with a specified radial distribution) 
and correlations between the galaxies. This approach applied to 
the galaxy-dark matter correlations
(which fully determines the g-g lensing signal)
was shown to give the same results as the simulations
in the regime of applicability 
\cite{2000MNRAS.318..203S}. 
However, halo model is analytical, does not suffer 
from the resolution issues and provides a more physical interpretation 
of the results. 
Thus one can parametrize the model with the 
quantities one wishes to extract from the data
and determine these directly.
It also allows for 
a rapid exploration of the parameter space without
the need to rerun cosmological simulations or to repopulate the halos with 
galaxies using semi-analytic galaxy formation models.
In this sense the halo model provides a 
natural link between the observations and the
theoretical models of galaxy formation.

Even though we will use the halo model to analyze g-g lensing the 
main features can be understood without it. Particularly robust 
conclusions are possible using the low density sample, where the 
clustering and group/cluster contribution can be neglected. In this 
case the signal can be taken simply as a projected radial mass 
profile of the 
halo averaged over the mass distribution of the halos determined by 
the galaxy sample. For the full and high density sample the 
contribution from groups and clusters can no longer be neglected, 
as it dominates the signal on scales above 200$h^{-1}$kpc. This 
allows one to determine the fraction of galaxies in groups and 
clusters as well. 

The outline of this paper is as follows: in section \S2 we review the 
basic theory and halo model as applied to g-g lensing. In \S3 we 
discuss the influence of the parameters introduced on the 
observed g-g lensing signal. We focus on the relative 
contributions to g-g lensing from galaxy, group/cluster and clustering 
terms and explore 
which parameters the observations are most sensitive to. In \S4 
we apply the model to the data to determine several of the model 
parameters. Interpretation of the results and 
conclusions are presented in \S5. 

\section{Contributions to g-g lensing from the halo model}

In the halo model it is assumed the galaxies form in collapsed
dark matter halos. The halos are clustered among themselves and have 
a specified density profile. 
The contribution to g-g lensing 
comes either from the dark matter profile of the halo the galaxy is 
sitting in (one-halo or Poisson term) or from the clustering of all 
the other halos around a given galaxy (halo-halo term). As we show below
the second contribution is small compared to the first one on scales below 
1$h^{-1}$Mpc of interest here. 
The one-halo term has two contributions. First is from the dark matter 
around the galaxy itself that depends on the dark matter halo profile. 
Second is from the dark matter in 
groups and clusters the galaxy may belong to, but is not at their center. 
In the latter case 
one must also
specify the radial distribution of galaxies inside these halos.
The simplest case is to assume it
is the same as the radial distribution of the dark matter 
(see e.g. \citeNP{2001astro.ph..9445H} for some justification in 
the case of observations and \shortciteNP{2000ApJ...544..616G}, 
\shortciteNP{2001MNRAS.328..726S} for the case 
of simulations). 

The relative contribution from the two one-halo components 
depends on the fraction of galaxies residing in groups and clusters. 
In general this fraction  
varies with luminosity, type and other selection criteria. 
Particularly simple is the low density sample in SDSS data, 
which consists mostly 
of the field galaxies and where the group/cluster
fraction is likely to be 
small. When this is not the case, as for example 
for the high density selected sample or early type sample, 
one must also address the question 
of whether the halo profiles around  
galaxies that are within larger 
halos differ from the profile of equal luminosity galaxies 
in the field. Even if the galaxies 
that end up in groups and clusters do not have different star formation 
efficiencies, so that their luminosity is a good indication of 
the halo mass, they may still loose some dark matter 
after merging. 
This depends on the amout of tidal stripping, which itself depends 
on the radial orbit of the galaxy.
Simulations indicate that the tidal radius should be defined at the 
orbit pericenter, which is the point of the closest approach to the center 
of the cluster \cite{2000ApJ...544..616G}. Since the orbits are very eccentric (typical ratios 1:5)
the pericenter can be significantly smaller than the instantaneous galaxy 
position. Still, 
these simulations indicate that the halo profiles of 
most of the halos within halos 
remain 
unaffected out to where the density of the larger halo matches that of the 
smaller halo \cite{2000ApJ...544..616G}. 
Since it is the radial halo profile that is relevant for g-g lensing
we will model the 
galaxies within clusters as a superposition of a smooth cluster 
profile (weighted by the radial distribution of galaxies) 
and an individual galactic halo profile, which is assumed to be the 
same as for the equal luminosity galaxy in the field. 
We discuss further below how this assumption can be tested with g-g lensing 
itself.
We do not need to model the transition from the subhalo to the cluster, 
where there may be 
departures from this simple model, 
since one is observing a projected density with g-g lensing.
Given that clusters are significantly larger than galactic halos
the cluster dominates the signal at radii smaller than the 3-d 
transition point. 

\subsection{Theory}

We are interested 
in the distribution of the dark matter around the galaxies
of a selected type, averaged over all the galaxies in the sample.
We can quantify this as an excess of the dark matter density above the
average as
a function of the radial separation $r$.
This is described by the galaxy-dark matter
cross-correlation  function,
\begin{equation}
\xi_{\rm g,dm}(r)=\left< \dd_{\rm g}(\vec{x})
\dd^{*}_{\rm dm}(\vec{x}+\vec{r}) \right>.
\end{equation}
Here $\dd_{\rm g}$ and	$\dd_{\rm dm}$ are the overdensities of
galaxies and dark matter, respectively. 

Observationally the quantity we are interested in is tangential shear
$\gamma_t$, which describes
elongation of images perpendicularly to the line connecting the
image and the lens. This is related to the convergence $\kappa=
\Sigma/\Sigma_{\rm crit}$, where $\Sigma$ is the projected surface 
density in units of critical density, 
\begin{equation}
\Sigma_{\rm crit}\gamma_t(R)\equiv \Delta \Sigma(R)=\bar{\Sigma}(R)-\Sigma(R) .
\label{dsigma}
\end{equation}
Here $R$ is the radial distance from the galaxy and $\bar{\Sigma}(R)$ is the 
mean surface density within $R$. $\Delta\Sigma(R)$
is the quantity that is observable, so 
all of our predictions will be compared to it. 
Critical surface density depends on the
distances to the lens and source galaxies, 
\begin{equation}
\Sigma_{\rm crit}={c^2 \over 4\pi G} {D_S \over D_L D_{LS}},
\end{equation}
where $D_L$ and $D_S$ are the angular diameter distances to the 
lens and source, respectively, and $D_{LS}$ is the angular diameter
distance between the two. In the case of SDSS we know the redshift of
the lens, so $D_L$ is known, while $D_S$ and $D_{LS}$ can only be determined 
in an average sense. 
Since SDSS survey is so shallow,
the redshift distribution of background galaxies is well determined from 
other deeper redshift surveys such as CNOC \shortcite{1995ApJ...455...50L}. This means that there 
is little error associated with the redshift distribution of background
population, which is not the case for deeper lensing surveys. 
While there is some uncertainty in the transformation from redshifts 
to distances
this is generally a small effect given the low redshifts of background 
population ($z \sim 0.3-0.5$). Here we will ignore it, assuming distances
using cosmological constant model with matter density $\Omega_m=0.3$.

To relate the observations to theory we interpret the above expressions in 
an average sense by averaging over all the galaxies and their surrounding 
dark matter. 
In this case the surface density $\Sigma$ is related to 
the galaxy-dark matter correlation function \cite{2001MNRAS.321..439G}
\be
\Sigma(R)=\int \bar{\rho} \xi_{\rm g,dm}[(R^2+\chi^2)^{1/2}) d\chi,
\label{sig}
\ee
where we dropped the unobservable constant term. Since we are interested
in the small scales where galaxies do not evolve much we will 
express everything in proper coordinates.
The low mean redshift of lens population ($\bar{z}\sim 0.1$) 
and the fact that most of the galactic halos have formed at higher 
redshift
means that the redshift 
evolution effects will be small if expressed in proper coordinates and
relative to the matter density today. We are implicitly 
assuming that most of the signal is arising near the galaxy, so we 
ignore the changes in the focusing strength along the line of sight.

\subsection{Halo model}
For a given correlation function $\xi_{\rm g,dm}$
we can use equations (\ref{dsigma}-\ref{sig})
to calculate $\Delta\Sigma$. In the halo model it is 
easier to calculate the power spectrum 
$P_{\rm g,dm}(k)$, which is just the Fourier 
transform of the correlation function. 
The halo model for power spectrum assumes the matter is in a
form of isolated halos with a well defined mass $M$ and halo profile
$\rho(r,M)$. The latter is defined to be an average over all
halos of a given mass and does not assume all halos have
the same profile.
The mass is determined by the total mass within the virial
radius $r_{\rm vir}$,
defined to be the radius where the mean 
density within it is $\delta_{\rm vir}$ times
the critical density of the universe. Here we will use the value 
$\delta_{\rm vir}=200$ relative to the critical density today. 
Often one uses the value defined 
by the spherical collapse model, which gives 
$\delta_{\rm vir} \sim 100$ for $\Lambda CDM$ model with 
$\Omega_m=0.3$. 
The latter gives about 15\% higher virial mass than the values 
presented here for a typical halo profile. 

We model the halo density profile in
the form
\begin{equation}
\rho(r)={\rho_s \over (r/r_s)^{-\alpha}(1+r/r_s)^{3+\alpha}}.
\label{rho}
\end{equation}
This model assumes that the profile shape is
universal in units of scale radius $r_s$, while its characteristic density
$\rho_s$ at $r_s$ or concentration $c=r_v/r_s$ may depend on the halo mass.
The halo profile is assumed to scale as
$r^{-3}$ in the outer parts and as $r^{\alpha}$
in the inner parts, with the transition between the two at $r_s$.
We fix the inner slope to
$\alpha=-1$ \cite{1997ApJ...490..493N}, since SDSS g-g lensing is not sensitive 
to small scales given that the first bin 
is at $R=75h^{-1}$kpc. 
For this profile the typical concentration on galactic scales is 
expected to be $c_{200}\sim 10$ 
(\shortciteNP{2001MNRAS.321..559B}, \citeNP{2001ApJ...554..114E}). 
We explore the sensitivity 
of g-g lensing to this parameter below. 

The first contribution to the one-halo or Poisson term comes from the
dark matter halos attached to the galaxy, where the galaxy is assumed 
to be at the halo center. 
If all the galaxies were
of the same mass then this term would be just the Fourier transform 
of the dark matter profile itself. A generalization of this is to 
integrate over the luminosity distribution assuming a relation 
between the galaxy luminosity $L$ and halo mass $M(L)$, 
\begin{equation}
P^{\rm cg}_{\rm g,dm}(k)= {1 \over (2\pi)^3 n}
\int {dn \over dL}
y[k,M(L)]dL,
\label{cp}
\end{equation}
where $dn/dL$ is the known luminosity distribution of the lens sample and 
$n$ is the total number of galaxies in the sample.
The superscript cg stands for central galaxy.
Here $y(k,M)=\tilde{\rho}(k,M)/M$, where $\tilde{\rho}(k,M)$ is the Fourier transform of
the radial halo profile of dark matter
\begin{equation}
\tilde{\rho}(k,M) = \int 4\pi r^2 dr \rho(r,M){\sin(kr) \over kr}.
\end{equation}
Note that unlike in previous work on halo model 
we do not require the profile to stop at 
the virial radius, so there can be more mass associated with the halo 
than the virial mass itself. This is only possible if we do not require 
the mass function defined below 
to integrate to unity, since some of the mass can be 
part of the extended halo structure beyond the virial radius. 
This is the case for the mass function by 
\shortciteN{2001MNRAS.321..372J} that we use here.
Typically we extend the halo to 2-3 times the virial radius and we 
verified that the exact cutoff position 
makes very little difference in the final results. 
As discussed above 
we assume the dark matter profile for a given luminosity does not 
depend on whether a galaxy is in a larger halo or not.
In the extreme case that the galaxies 
within larger halos do not have any mass attached to them at all the 
virial masses of field galaxies should be underestimated by the fraction 
of galaxies in groups and clusters. 
Most of these galaxies are in fact in the field, so this fraction and 
the corresponding correction is small.

Second contribution to the one-halo term comes from the
dark matter around galaxies in groups and clusters, where 
the dark matter is not associated with the galaxy but with a
larger halo. 
This term by definition includes only the galaxies that are not 
at the center of the halo they find themselves in, since such
galaxies are already accounted for by the previous term. 
To obtain the g-g signal one must average over the
mass distribution of halos and over the distribution of galaxies 
within these halos. 
Since the halos come in a variety of sizes we must first introduce the halo mass 
function $dn / dM$, describing the number density of halos
as a function of mass. It can be written as
\begin{equation}
{dn \over d\ln M} ={\bar{\rho} \over M}f(\sigma){d \ln \sigma^{-1} \over
d \ln M},
\end{equation}
where $\bar{\rho}$ is the mean matter density of the universe, $M$ is the
virial mass of the halo and $n(M)$ is the spatial number density of halos
of a given mass $M$. We
introduced a
function $f(\sigma)$, which
has a universal form independent of the
power spectrum, matter density, normalization or redshift if written
as a function of rms variance of linear density field $\sigma$.
\shortciteN{2001MNRAS.321..372J}
propose the following form (see also \citeNP{1999MNRAS.308..119S}),
\begin{equation}
f(\sigma)=0.315 \exp[-|\ln \sigma^{-1}+0.61|^{3.8}],
\end{equation}
which they argue is universal if mass used is within the
radius where overdensity is 200 of mean density (since we use 
a different definition of halo virial mass we must correct for the  
difference between the two). Note that this mass function does 
not account for all the mass density in the universe, 
so that some of the mass could be present 
in the outer parts of the halos. 

\begin{figure}
\begin{center}
\leavevmode\epsfxsize=8cm \epsfbox{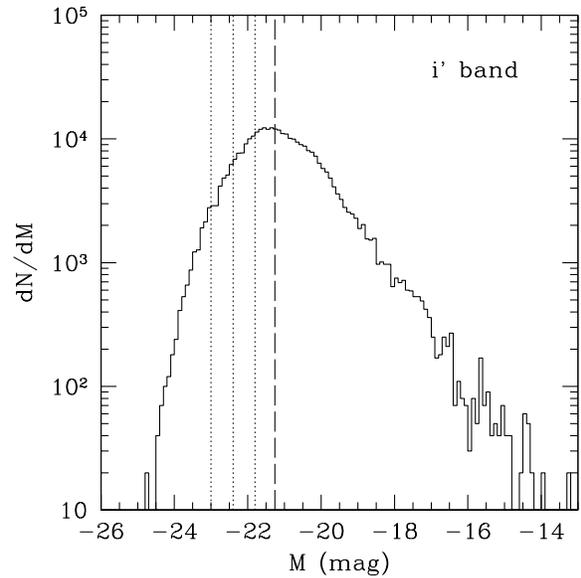}
\end{center}
\caption{Magnitude
distribution of the SDSS sample in $i'$ band (in units with $h=1$) for the 
whole sample and the 4 luminosity subsamples separated by dotted lines. 
Also indicated is the 
$L_{\star}$ magnitude \protect\shortcite{2001AJ....121.2358B} 
as a vertical dashed line, which roughly corresponds to the 
mean luminosity of the whole sample ($\langle L \rangle\sim L_{\star}=2.05\times 
10^{10}h^{-2}L_{\sun}$). Note that even though the mean luminosity is close 
to $L_{\star}$ most of the lensing signal comes from the massive
galaxies  
above $L_{\star}$, which explains the choice of luminosity subsamples.}
\label{fig1}
\end{figure}

We must still specify how the galaxies populate the halos of different 
mass. For any galaxy sample this can be parametrized as
the mean number of galaxies as a function of halo mass 
$\langle N \rangle (M)$ \cite{1998ApJ...494....1J}.
Here we work with $\langle N\rangle(M)$ averaged over the 
luminosity distribution of the sample, which is dominated by 
the galaxies around $L_{\star}$ (see figure \ref{fig1}). We describe a 
model for $\langle N\rangle(M)$ below. The contribution from 
this term is 
\begin{equation}
P^{\rm gc}_{\rm g,dm}(k)= {1 \over (2\pi)^3 \bar{n}}
\int f(\sigma)d\ln \sigma \langle N \rangle
y(k,M)y_g(k,M),
\label{gc}
\end{equation}
where $y_g$ is defined as the Fourier transform of the
radial distribution of galaxies and the superscript gc stands for group
and cluster contribution. 
We will assume $y_g=y$ in most of this paper, but we also explore the 
sensitivity to this assumption in the next section. 

Finally, the halo-halo contribution to the 
power spectrum describes the correlations between 
the lens galaxy and neighbouring halos. Here we assume it 
follows the linear power spectrum, except that halos 
can be biased relative to it, with bias $b(\sigma)$ a function of 
rms variance $\sigma$. This assumption has been shown to give 
a good agreement with simulations of galaxy-dark 
matter correlations \cite{2000MNRAS.318..203S}. 
Low mass halos are unbiased or mildly antibiased ($b \leq 1$, 
\citeNP{2001MNRAS.323....1S}). 
The power spectrum of this term 
can be calculated using \cite{2000MNRAS.318..203S}
\begin{eqnarray}
P^{hh}_{\rm g,dm}(k)&=&P_{\rm lin}(k) \left[{\bar{\rho}\over \bar{n}}
\int f(\sigma)d\ln \sigma{\langle N \rangle \over M}b(\sigma)y_g^p[k,M(\sigma)]
\right] \nonumber \\
&\times &
\left[\int f(\sigma)d\ln \sigma  b(\sigma)y[k,M(\sigma)]\right],
\label{chh}
\end{eqnarray}
where $p=0$ for the field component and $p=1$ for the group/cluster 
component. Here we
implicitly assumed that $\langle N \rangle(M) $ 
includes also the central galaxy component, which means it will have 
a strong peak at the halo mass corresponding to the $L_{\star}$ luminosity 
(figure \ref{fig1}). Since we will show that 
this term is negligible 
on the scales below $1h^{-1}$Mpc of interest here the details of its modeling are not
very important for our purpose.

The total power spectrum is simply the sum of all contributions. 
Once this is computed we Fourier transform it to obtain the correlation 
function and integrate it along the line of sight (equation \ref{sig})
to obtain the projected surface density $\Sigma(R)$. Another integration 
(equation \ref{dsigma}) is needed to obtain the mean surface density 
within $R$ and the observed quantity  
$\Delta\Sigma(R)$. 
Since the redshift of lenses is low
we do not include any redshift evolution effects.

The above model left several functions unspecified, most notably 
the halo mass-luminosity relation and 
the galaxy occupation 
number $\langle N \rangle(M)$. 
We assume that 
there is a one to one correspondence between the halo mass and luminosity 
of the galaxy, which we model as a power law
\begin{equation}
{M \over M_{\star}}=\left({L \over L_{\star}}\right)^{\beta},
\label{ml}
\end{equation}
where $M_{\star}$ is the mass associated with $L_{\star}$ galaxy. The slope $\beta$
parametrizes the relation for a simple power law dependence. This is 
not the most general possibility and over a broad range of luminosity 
$\beta$ is expected to vary, but since we
are working over a narrow range of luminosities (a factor of 10 at best) 
this parametrization should not be too bad and
in any case, the data do not yet allow for a more 
general parametrization. Note that bright red galaxies 
have been removed from the 
sample, since it is likely that they do not fall on the same 
mass-luminosity relation. 
Both 
$M_{\star}$ and $\beta$ can be extracted from the data. We could
also give some scatter to this relation, but in practice the 
scatter is dominated by the fact that we cannot choose very narrow
luminosity bands for the analysis and so we will not pursue it 
further here. 
The distribution of halo 
masses has a width that depends on the 
luminosity distribution of the galaxies as shown in figure \ref{fig1}.

The second unspecified function is the 
galaxy occupation
number $\langle N \rangle(M)$ for noncentral galaxies in groups and clusters. 
This is expected to grow 
with the halo mass for any galaxy population. 
Here we assume that the mean number of galaxies 
(of a given type) has a power law relation to the mass of the halo, 
\begin{equation}
\langle N \rangle \propto M^{\epsilon}, \,\,\,\,M > M_{\rm cutoff}.
\label{nm}
\end{equation}
We assume the power law relation is valid from some minimum halo mass 
$M_{\rm cutoff}$, 
which should be at least a few times above the typical halo mass for a 
given luminosity sample, 
since by definition it is assumed these galaxies are not at the halo centers, 
where presumably another bright galaxy resides. We explore the 
sensitivity to the low mass cutoff $M_{\rm cutoff}$ and $\epsilon$ below. 
We normalize the group/cluster contribution in terms of a
fraction of galaxies $\alpha$ that reside in these larger halos. 

\begin{figure}
\begin{center}
\leavevmode
\epsfxsize=8.0cm \epsfbox{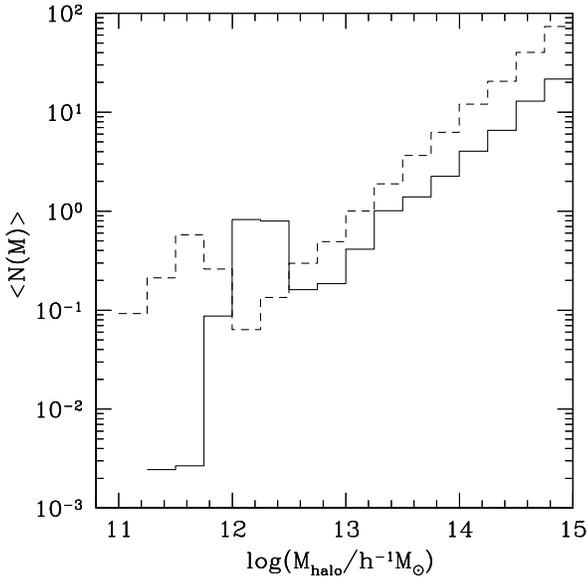}
\end{center}
\caption{
$N(M)$ calculated from SAMs using GIF simulations \protect\shortcite{1999MNRAS.303..188K}
using two narrow luminosity samples, $-20.5<I<-20$ (dashed)
and $-22<I<-21$ (solid). Both the central galaxy component peaking 
at low halo masses and non-central group/cluster 
component are included here. The latter has $N(M)$ 
approximately proportional to $M$, ie $\epsilon=1$.
}
\label{fig3}
\end{figure}

In figure \ref{fig3} we show $\langle N\rangle/M$ 
versus $M$ as calculated from semi-analytic  simulations 
\shortcite{1999MNRAS.303..188K} for two 
narrow intervals in luminosity, a brighter one 
around $L_{\star}$ and a fainter one magnitude below it. 
One can see that in both cases 
the distribution is well modeled as a sum of a narrowly peaked distribution for 
the central galaxies  
and a linear power law component $\langle N\rangle \propto M$ for the 
noncentral galaxies in groups and 
clusters. In our model we would predict
first contribution to be narrow because we have chosen a narrow bin in 
luminosity, which is strongly correlated with the halo mass for central galaxies.
This is clearly seen to be the case in figure \ref{fig3}. 
Second contribution is a power law because more massive halos also contain more 
subhalos of a given mass, which host noncentral galaxies.
It is the latter contribution that enters in our model for $\langle N\rangle/M$ 
in equation \ref{cp}, 
while the first component is included by the galaxy 
luminosity function.
Our model therefore qualitatively 
reproduces the main features of SAMs, but by  
characterizing the main ingredients with a few free parameters 
it allows us to be more general than any specific model. 

It is interesting to compare the fraction $\alpha$ 
of galaxies in groups/clusters 
for the two cases from figure \ref{fig3}. 
This is 20\% for the bright sample and 25\% for 
the faint sample. This parameter can therefore 
vary as a function of luminosity and is expected 
to decrease with $L$. This means that the group/cluster
contamination is less important for bright galaxies, where 
it becomes more difficult to determine it observationally because of the
small number statistics. We will use this fact in the modeling below. 
 
\section{Dependence of g-g lensing on the halo model parameters}

\subsection{Halo mass and concentration}

\begin{figure}
\begin{center}
\leavevmode
\epsfxsize=8.0cm \epsfbox{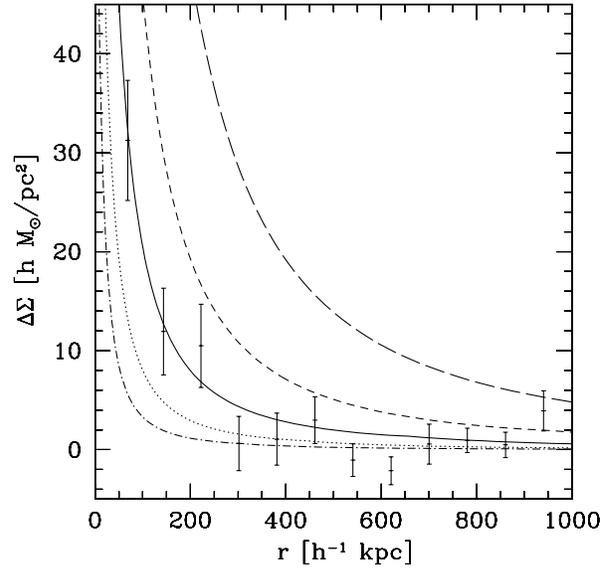}
\end{center}
\caption{
Signal for a NFW halo profile with $c=10$ 
of varying mass. From bottom to top 
$M=10^{11}, 3\times 10^{11}, 10^{12}, 3\times 10^{12}, 10^{13} h^{-1}
M_{\sun}$. Also shown are the observational data for the low density sample.
}
\label{fig4}
\end{figure}

We have introduced several free parameters in the model above and
in this section we explore the dependence of the measured signal 
on them.
The main parameter is the virial halo
mass $M_{\star}$ of a typical $L_{\star}$ galaxy. 
The resulting g-g lensing signal for halos of masses between 
$10^{11}-10^{13}h^{-1}M_{\sun}$ is shown in figure \ref{fig4}. 
For reference we also show the SDSS data using the low density 
subsample, described in more detail below.
Only the one-halo term is shown and we used NFW profile with 
$c=10$ and a $\delta$-function in the halo mass distribution. 
What is shown is therefore a simple projection of NFW profile, 
which has a well known analytic form for $\Sigma(R)$ \cite{2001PhR...340..291B}.
The signal at a given scale is rapidly increasing with 
the mass of the halo, indicating that 
SDSS g-g lensing is a very sensitive probe of the 
virial mass of the halo. Masses around 
$10^{12}h^{-1}M_{\sun}$ 
seem to be required to explain the field sample. 

\begin{figure}
\begin{center}
\leavevmode
\epsfxsize=8.0cm \epsfbox{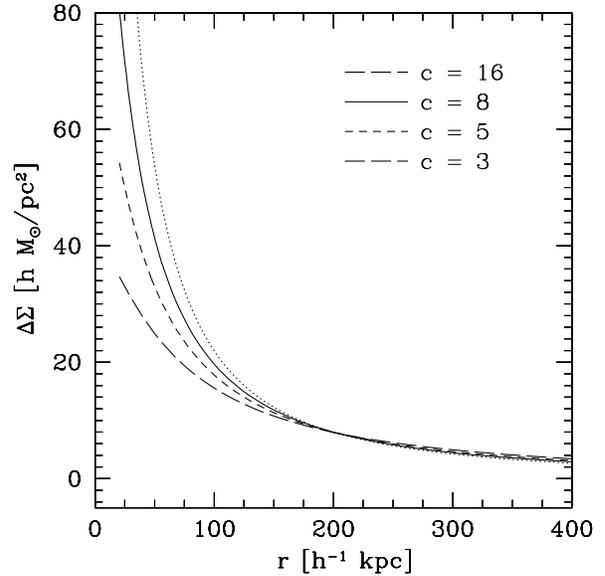}
\end{center}
\caption{
Signal for a halo with NFW profile of fixed mass varying 
concentration $c$ from 3 to 16. 
}
\label{fig5}
\end{figure}

The dependence on the concentration parameter $c$ is shown in 
figure \ref{fig5}. We keep the virial mass of the halo constant, 
so the signal at large radii is similar for all the models. 
At smaller radii the profiles start to differ and the more 
concentrated halos give a larger signal than the less concentrated ones.
Doubling the concentration index $c$ changes the signal by 20-30\%
in the inner most bin at 75$h^{-1}$kpc and less than that 
at larger radii.
We will see below that 
since some contribution to the mass determination 
comes from the inner most bin there is some degeneracy between $c$
and $M$, such that larger values for $c$ give lower values for $M$. 
However, this degeneracy is not complete and
too steep profiles can be ruled out regardless of the halo mass. 

\begin{figure}
\begin{center}
\leavevmode
\epsfxsize=8.0cm \epsfbox{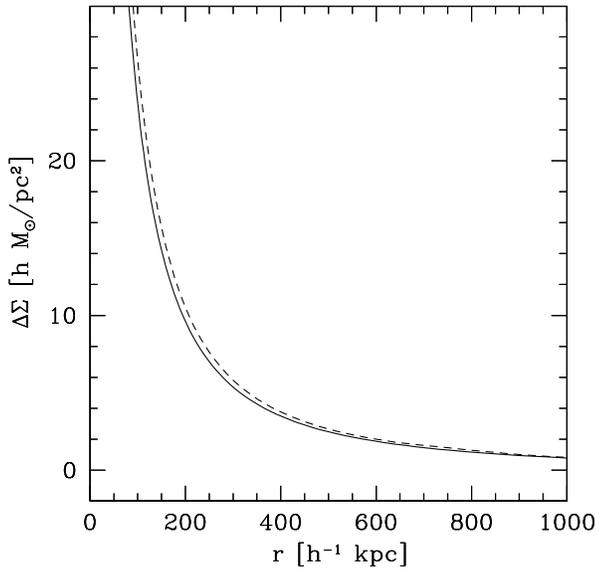}
\end{center}
\caption{
Comparison between a single NFW profile for $M_{\star}$ (solid) and an average 
profile using the realistic luminosity distribution and assuming 
luminosity is proportional to mass (dashed).} 
\label{fig6}
\end{figure}

Since we cannot choose an infinitely narrow distribution in 
luminosity and have a high signal to noise 
a $\delta$-function in the halo mass distribution is 
not very realistic. Instead one must integrate over the halo 
mass distribution to obtain the average profile. While the mean 
signal will correspond to the mean halo mass, the 
shape of the signal may be more affected, possibly 
complicating the determination of the concentration parameter. To 
investigate this possibility 
we compute the signal for the luminosity distribution 
in figure \ref{fig1}, assuming halo mass is proportional to luminosity and
comparing it to a single NFW profile corresponding to the mean mass of 
the sample. The result is shown in figure \ref{fig6}. The two profiles
are very similar, indicating that even for realistic luminosity distributions
the shape of the halo can be reliably determined from the data if all 
the galaxies are at the halo centers.
These results show that if the galaxies are in the field then g-g 
lensing is a very robust probe of the halo virial mass and its profile. 
The field assumption should be valid for the low density sample 
discussed in more detail below. For the whole sample one must introduce 
also the contribution from groups and clusters, which is discussed next. 

\subsection{Group and cluster contribution}

\begin{figure}
\begin{center}
\leavevmode
\epsfxsize=8.0cm \epsfbox{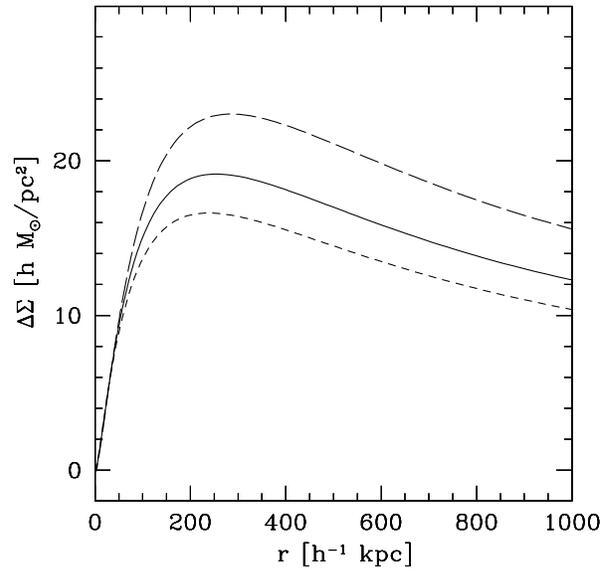}
\end{center}
\caption{
Group and cluster contribution to g-g lensing as a function of lower mass 
cutoff $M_{\rm cutoff}$. 
Results shown here are for $M_{\rm cutoff}$
of $10^{12}h^{-1}M_{\sun}$ (bottom curve), 
$3\times 10^{12}h^{-1}M_{\sun}$ (middle curve) and $ 10^{13}h^{-1}M_{\sun}$ 
(top curve). We use $c=7$ for galaxy and dark matter halos.
}
\label{fig7}
\end{figure}

The group and cluster contribution to lensing 
is shown in figure \ref{fig7} 
for $\epsilon=1$. Here it is assumed
all the galaxies are in groups and clusters ($\alpha=1$).
It is evident that this contribution is significantly 
different from the central galaxy contribution. At small radii the 
group/cluster contribution is 
negligible, since by definition these galaxies are not at the halo center 
(the 
individual subhalo contribution from these noncentral galaxies is already 
included in the previous term), 
the cluster component itself is very smooth and the galaxies are assumed
to be distributed radially 
like dark matter and do not have a strong central peak. We remind the 
reader that for a uniform density the shear signal goes to zero 
by equation \ref{dsigma}.
The group/cluster contribution peaks around
200$h^{-1}$kpc and then gradually decreases with radius, such that even at 
1$h^{-1}$Mpc it is still 60\% of the peak value. 
One can see that varying the low mass cutoff $M_{\rm cutoff}$
has some effect on the amplitude of the signal. For lower
mass cutoffs one finds a lower average signal, because there are 
proportionally more galaxies in less massive halos. 
On the other hand, the 
radial dependence of this contribution is less affected. 
Since the radial dependence of this contribution 
is very different from the central galaxy contribution (figure \ref{fig4})
the two contributions can be robustly separated in the data even if the 
actual value of $\alpha$ has some uncertainty related to the low mass cutoff.
Based on the results in figure \ref{fig3} we choose $M_{\rm cutoff}$ to 
be 3 times the halo mass corresponding to the luminosity of the sample.
For $L_{\star}$ galaxies this is around
$2\times 10^{12}h^{-1}M_{\sun}$. 
This cutoff is increased when we analyze brighter luminosity subsamples.

\begin{figure}
\begin{center}
\leavevmode
\epsfxsize=8.0cm \epsfbox{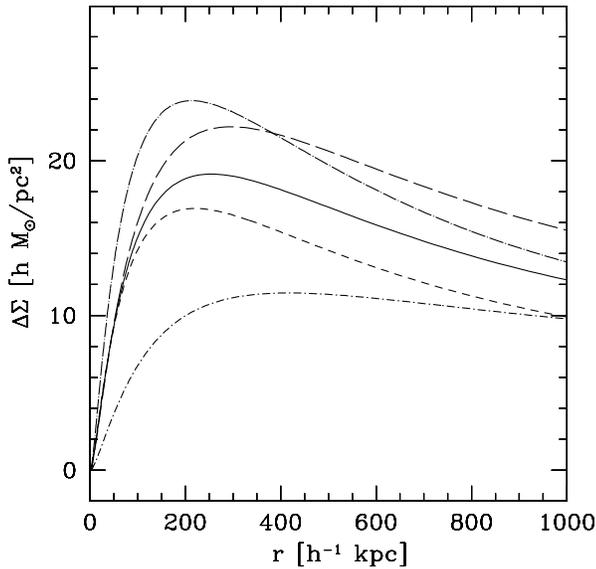}
\end{center}
\caption{
Group and cluster contribution to g-g lensing as a function of 
$\epsilon$=0.8, 1.0, 
1.2 (short dashed, solid, long dashed, respectively) and using 
$c=7$ for galaxy and dark matter. Also shown 
is variation of radial galaxy distribution
parametrized with concentration $c=3$ 
(dot-short dashed) and $c=10$ (dot-long dashed) for $\epsilon=1$.
}
\label{fig8}
\end{figure}

In addition to the group low mass cutoff we 
also explored the sensitivity of this component to 
the slope $\epsilon$ of the galaxy occupation number (equation \ref{nm}) 
and to the radial galaxy distribution $y_g$ (equation \ref{cp}).
The results are shown in figure \ref{fig8}. 
Higher $\epsilon$ gives more weight to more massive clusters and 
has a higher signal than a lower $\epsilon$. Similarly, more concentrated 
galaxy distribution enhances the signal since now more galaxies are 
closer to the cluster center where the halo density is higher.
Most observations 
and simulations predict that galaxy distribution does not deviate 
significantly from the dark matter, which has a mean concentration 
of around $c=5-10$ for groups and clusters, depending on the low mass cutoff.  
Here we compare the cases with $c=3,7,10$ for the galaxy while using 
$c=7$ for the dark matter. 
In all cases the overall amplitude changes somewhat depending on the 
parameters, while the qualitative 
shape of the curve remains similar, peaking 
around 200$h^{-1}$kpc and then decreasing with radius. 
This is distinguishable from the central galaxy component regardless of the 
specific choice of the parameters. In the following we
assume $\epsilon=1$ and $y_g=y$ with $c=7$. 

\begin{figure}
\begin{center}
\leavevmode
\epsfxsize=8.0cm \epsfbox{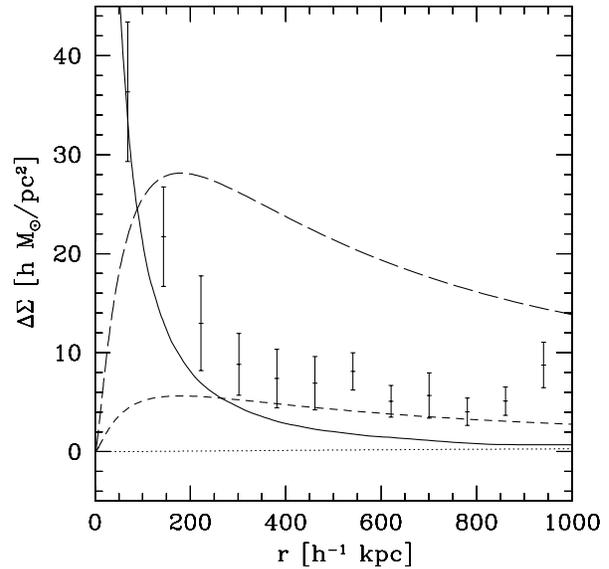}
\end{center}
\caption{Central galaxy contribution for $M_{\star}=10^{12}h^{-1}M_{\sun}$ (solid), 
group/cluster contribution 
with $\alpha=0.2$ (short dashed) and $\alpha=1$ (long dashed) and halo-halo 
contribution to g-g lensing (dotted). Also shown is the SDSS data for the 
whole sample. } 
\label{fig9}
\end{figure}

To combine the central galaxy and group and cluster
contributions we must assume parameter $\alpha$, which 
is the fraction of galaxies in groups and clusters. Variation in this 
parameter is shown in figure \ref{fig9}. If the fraction $\alpha$
is small the signal on large scales is also small, whereas
in the opposite case the large scale and small scale signals are 
comparable. 
Also shown are the data from the whole SDSS sample, which show a relatively flat 
signal between 300-1000$h^{-1}$kpc. This cannot be explained by the central 
galaxy contribution alone, which drops below the observed points at $R>200h^{-1}$kpc.
It also cannot be explained by all the galaxies being in groups and clusters,
which would have an excess of signal at large radii.
A contribution from noncentral galaxies in 
groups and clusters with $\alpha=0.2$ 
together with the central galaxy 
contribution provides a good description of the data.

The contribution from the correlations between the galaxies and the neighbouring halos
(halo-halo term) is also shown in figure \ref{fig9}. 
This contribution is 
very small and cannot explain the strength of the signal on scales around 
300-1000$h^{-1}$kpc. Instead the group and cluster contribution is 
required to explain the signal 
on these scales. Note that
this does not mean that the correlations 
between the galaxies are not important, since the galaxies residing in groups and 
clusters are also correlated among themselves. 
It does however mean that the correlations 
between the galaxies in isolated halos are small and are instead
dominated by the galaxies that reside in groups and clusters, whose 
contribution to g-g lensing is included in group and cluster term
discussed above.  
The main difference between the two terms is that for the latter most of 
the mass in large halos is not associated with the halos of 
individual galaxies within them, 
which account for only about 5-10\% of the mass, 
but with a diffuse component of groups or clusters. In this 
respect our analysis differs from the previous analysis of this data 
\shortcite{2001astro.ph..8013M},
where only the mass associated with the visible galaxies was
accounted for in the correlation analysis. 

\section{Application to the SDSS data}

In this section we apply the formalism developed in previous sections 
to the SDSS data. We begin first by analyzing the luminosity dependence of 
g-g lensing, which provides us with a parametrization of the mass luminosity 
relation and with a typical mass $M_{\star}$ of an $L_{\star}$ galaxy. 
We proceed by analyzing the full sample as well as high and low 
density versions of it, which provide an alternative
estimate of $M_{\star}$ and the fraction $\alpha$ of group/cluster 
galaxies in the sample, as well as an estimate of mass loss in 
dense enviroments. Finally, we address the morphological dependence
of the signal, which helps to explain some of the trends in mass-luminosity
dependence. 

\subsection{Luminosity dependence}

In this subsection we model the luminosity dependence of the signal using 
the g-g lensing as a function of galaxy luminosity in all 5 SDSS colors. 
The lenses have been divided into 4 luminosity bins, with the mean luminosity 
varying between
0.6$L_{\star}$ to 7$L_{\star}$ (see figure 
\ref{fig1}). 
In all cases the quantity we analyze is $\Delta \Sigma(R)$, 
averaged in equally spaced radial bins. 
We assume that the errors between bins are independent. This is 
a good approximation at small separations, where a given background galaxy 
contributes typically only to one lens galaxy (average separation between 
the lens galaxies is around 300$h^{-1}$kpc). 
At larger separations this assumption 
breaks down, but since we will be mostly focusing on the inner
regions this will not be a significant factor for the results of the 
fits (although it may affect the goodness of fit, see below).

The data divided in luminosity bins do not have sufficient statistical 
signal to allow accurate determination of group/cluster fraction in each bin 
separately.
Since this fraction is expected to decrease with luminosity (figure \ref{fig3})
and it is already low in the lowest luminosity bin 
we make the fits using two different assumptions. In the first we 
assume $\alpha=0$ for all but the lowest luminosity bin. 
In the 
second we assume $\alpha$ from the lowest luminosity bin also applies to 
the higher luminosity bins and we use luminosity dependent $M_{\rm cutoff}$
roughly 3 times above the mass corresponding to that luminosity.
Since $\alpha$ is expected to decrease with luminosity the answers we 
get from the two models 
should bracket the real values.
We do not fit for the mean mass in each bin 
separately, but use the power law relaton between luminosity 
and mass to fit for 
$M_{\star}$ and $\beta$.

\begin{figure}
\begin{center}
\leavevmode
\epsfxsize=8.0cm \epsfbox{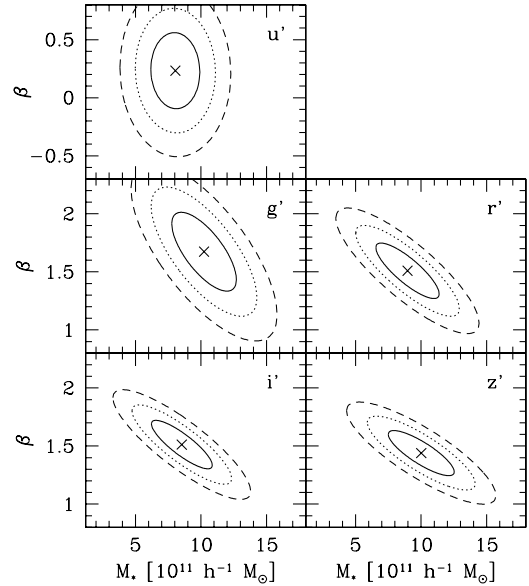}
\end{center}
\caption{
68\%, 95\% and 99\% contour plots for $\beta$ and 
$M_{\star}$ as a function of SDSS color for the case $\alpha=0$ in all 
but the lowest luminosity bin. The results are
consistent with $M_{\star} \sim 6-8\times 10^{11}h^{-1}M_{\sun}$ and 
$M/L \propto L^{1/2}$ except in $u'$ where no clear 
correlation between light and 
mass is detected.
}
\label{fig10a}
\end{figure}

\begin{figure}
\begin{center}
\leavevmode
\epsfxsize=8.0cm \epsfbox{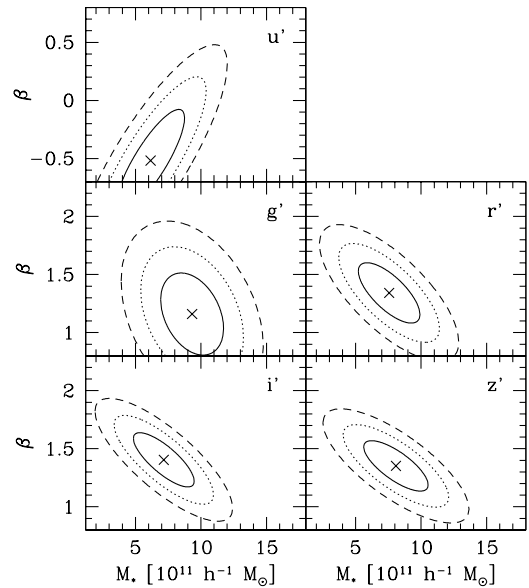}
\end{center}
\caption{
Same as \ref{fig10a} but 
with equal $\alpha$ in all 
luminosity bins. 
}
\label{fig10b}
\end{figure}

\begin{figure}
\begin{center}
\leavevmode
\epsfxsize=8.0cm \epsfbox{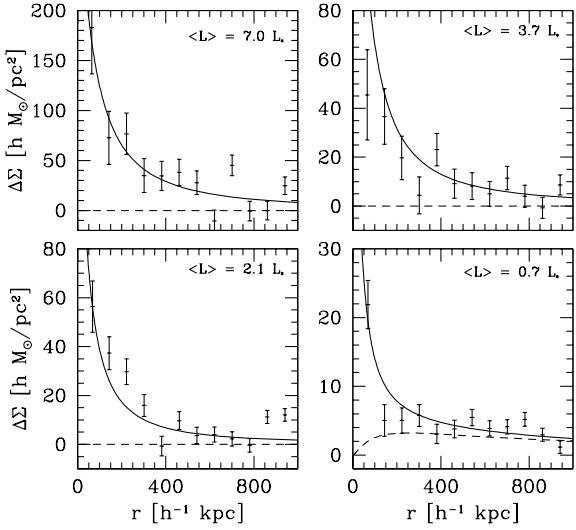}
\end{center}
\caption{Dependence of the g-g signal on luminosity using the 
4 subsamples in $i'$ (figure \ref{fig1}) together with the best fitted
model (solid). Dashed shows just the group and 
cluster contribution. 
We assume $\alpha=0$ except for the lowest luminosity bin.} 
\label{fig11a}
\end{figure}

\begin{figure}
\begin{center}
\leavevmode
\epsfxsize=8.0cm \epsfbox{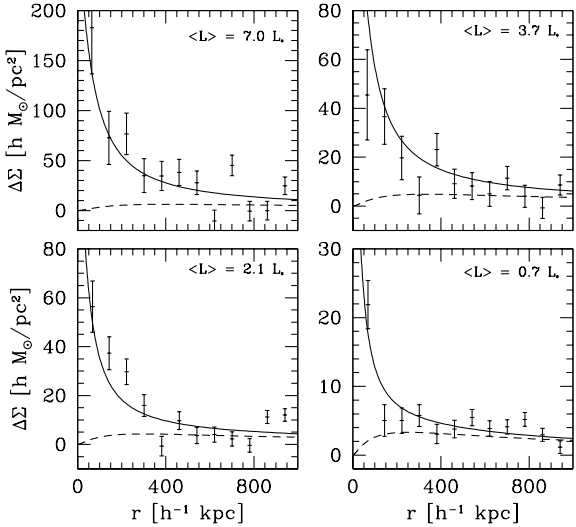}
\end{center}
\caption{Same as figure \ref{fig11a} but 
with same $\alpha$ in all 
luminosity bins.
} 
\label{fig11b}
\end{figure}

The results from the fits for both cases 
are given in table \ref{table1} and shown in figures \ref{fig11a} and 
\ref{fig11b}. 
The contour plots for $M_{\star}$ and $\beta$
are shown in figures \ref{fig10a},  \ref{fig10b} for all bands. 
As one can see from table \ref{table1} except for the bluest band $u'$
the relation between luminosity and mass is 
well established, as already shown by the SDSS team 
\cite{2001astro.ph..8013M}. The origin for lack of correlation in $u'$ is 
the morphology dependence of the signal, as discussed in more detail 
below. 
In red bands 
the best fitted value for $\beta$ is significantly 
above unity, 
indicating that the virial mass to light ratio is increasing with 
the halo mass (or luminosity). 
If light is a good tracer of stellar 
mass (as expected in the reddest bands) this indicates that 
the star formation efficiency is decreasing with luminosity. 
For example, in $i'$ the fitted values are 
$M_{\star}=(6\pm 2)\times 10^{11}h^{-1}M_{\sun}$, $\alpha=0.2\pm 0.05$ 
and $\beta=1.51\pm 0.15$, so that the value $\beta=1$ is more than 
3-$\sigma$ away from the best fitted value. This gives 
$M_{200}/L\sim 30hM_{\sun}/L_{\sun}$ at $L_{\star}$, which is the lowest 
luminosity that still has a detectable signal.
This increases to
$M_{200}/L\sim 80hM_{\sun}/L_{\sun}$ at $7L_{\star}$ corresponding to the 
brightest luminosity bin.
If instead $M_{100}$ is used as the virial mass 
then these values should be increased
by an additional 15\%. These results however still hide significant
variations between morphological types, as discussed below. 
Note that there is a strong correlation between 
$M_{\star}$ and $\beta$. This is because even though $L_{\star}$ is close 
to the mean luminosity of the sample, the luminosity distribution is 
broad and most of the signal comes from the brighter end of the 
luminosity distribution. An increase in $\beta$ gives more mass to brighter 
galaxies, which allows for a lower $M_{\star}$ to fit the data.

\begin{figure}
\begin{center}
\leavevmode
\epsfxsize=8.0cm \epsfbox{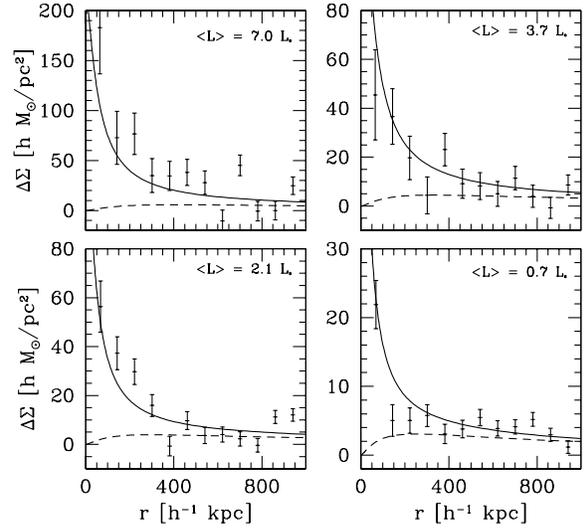}
\end{center}
\caption{Same as figure \ref{fig11b} but with $\beta=1$.
} 
\label{fig11c}
\end{figure}

In figure \ref{fig11a} 
the best fitted model is compared to the data in $i'$ band for the
case where $\alpha=0$ except in the lowest luminosity bin. 
The model provides a reasonable fit to the data. 
The reduced $\chi^2$ is around 2.4, indicating that 
the fit is not perfect, although there is no 
obvious systematic deviation that would indicate a clear 
shortfall of the model, suggesting instead that the observational 
error bars may be somewhat underestimated. 
One explanation for this is neglection of
correlations between the bins, although this should only be 
important at large radii.
Figure \ref{fig11c} shows the best fitted model assuming $\beta=1$. 
For this case we find $M_{\star}=1.5\times 10^{12}h^{-1}M_{\sun}$ 
in $i'$, but the fit is less good now. 

For the case with the same value of $\alpha$ for all 4 bins we typically find 
$M_{\star}$ is about 15\% lower as shown in figure \ref{fig10b} and in table 
\ref{table1}, 
because part of the signal in the high 
luminosity bins is now accounted for by the group/cluster contribution.
The slope $\beta$ is reduced as well. 
The reduced $\chi^2$ reduces to 2, so this procedure does somewhat better 
in describing the data, as shown in figure \ref{fig11b}. 
The value for $\beta$ does not change 
significantly (figure \ref{fig10b}).

\subsection{Density selection and the fraction of group/cluster galaxies}

\begin{figure}
\begin{center}
\leavevmode
\epsfxsize=8.0cm \epsfbox{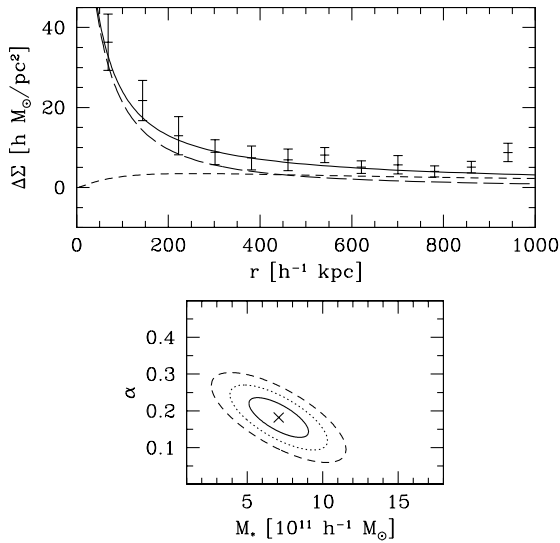}
\end{center}
\caption{Top: best fitted model (solid) to the complete sample with parameter 
values given in the text. Also shown are central galaxy contribution 
(long dashed) and group/cluster contribution (short dashed). We do not 
show the clustering (halo-halo) contribution, which is negligible.
Bottom: contour plot of $\alpha$ and $M_{\star}$. We used luminosity 
distribution in $i'$ for this plot. 
}
\label{fig12}
\end{figure}

In previous subsection we analyzed the luminosity dependent data to 
obtain the relation between the mass and luminosity around $L_{\star}$ galaxies. 
In this section we combine the luminosity data 
into a complete sample of galaxies
\footnote{Note that averaging the luminosity bins does not give the same result
as the average given in \shortciteN{2001astro.ph..8013M}. 
The reason for the difference is the clustering 
correction applied to the data. Since some fraction of background galaxies
is correlated with the lens they do not contribute to the lensing, 
so one must correct upwards the measured signal. This correction 
increases with the luminosity of the lens galaxy and the proper procedure
is to apply the correction to the luminosity dependent sample and then 
average the luminosity bins. 
Future data with better photometric redshift information for 
the background galaxies may allow one to eliminate this procedure by 
selecting only background galaxies that are far from the lens.}.
This allows us to 
determine more accurately the fraction of galaxies in groups and clusters, 
which could not be modeled for each luminosity bin separately. 
Using $\beta$  
from the previous subsection and the actual luminosity distribution 
(figure \ref{fig1})
we fit for $M_{\star}$ and $\alpha$. In $i'$ we find 
$M_{\star}=(6 \pm 1)\times 10^{11}h^{-1}M_{\sun}$ and 
$\alpha=0.19 \pm 0.03$. These values are consistent with the values
obtained in previous subsection for the case with $\alpha$ 
independent of luminosity.
The resulting fits are shown in figure \ref{fig12}, where we see that the 
model provides a good description of the data (reduced 
$\chi^2 \sim 1$).
It is clear that both the 
central galaxy and the group/cluster components are needed to explain 
the data.

\begin{figure*}
\begin{center}
\leavevmode
\epsfxsize=10.0cm \epsfbox{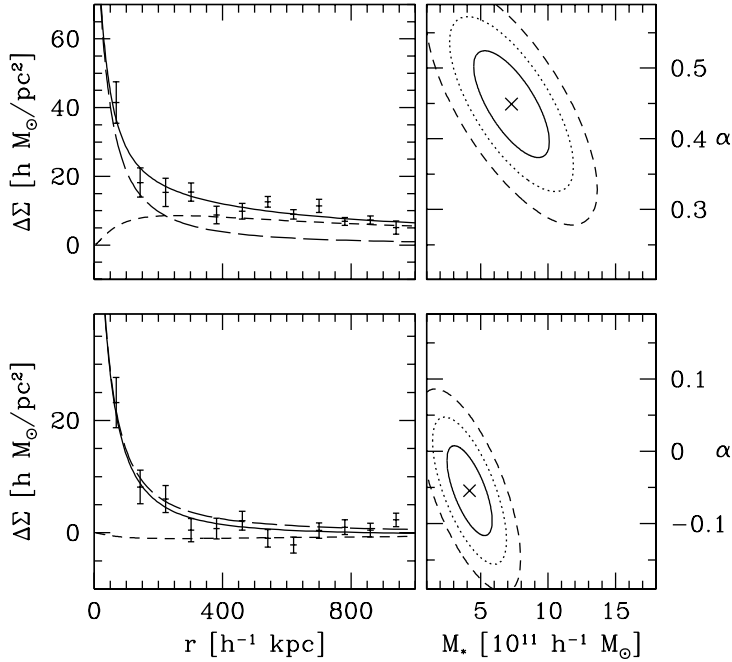}
\end{center}
\caption{Top: best fitted model to the high density sample together 
with the contour plot in $M_{\star}-\beta$ plane for luminosity 
distribution in $i'$. 
Bottom: same for the low density sample. 
}
\label{fig13}
\end{figure*}

To further test the obtained values we use the data split into
high and low density enviroments \shortcite{2001astro.ph..8013M}. 
The separation is done using 
Voronoi tesselation, which assigns a weight to each galaxy according 
to the local density of galaxies. The galaxies are then rank ordered 
and divided into two equal samples. Both samples have roughly equal 
average luminosity, so under the assumption that luminosity scales with
mass independent of enviroment
the estimated $M_{\star}$ should be approximately 
equal. 
The resulting fits are shown in figure \ref{fig13}.
For the low density sample the best fit gives 
slightly negative $\alpha=-0.07$, showing that 
the low density sample is consistent with the 
galaxies being only in the field.
The best fitted mass in this case 
is $M_{\star}=(6-7)\times 10^{11}h^{-1}M_{\sun}$
depending on the passband. 
If we assume $\alpha=0$ then we find
$M_{\star}=(4.5 \pm 1.5)\times 10^{11}h^{-1}M_{\sun}$ for the low density sample 
in $i'$. For the
high density sample the values are 
$M_{\star}=(5-8 )\times 10^{11}h^{-1}M_{\sun}$ and
$\alpha=0.45 \pm 0.04$. Assuming $\alpha=0$ for low 
density sample one would expect $\alpha\sim 2\times0.2=0.4$ for the high 
density sample using $\alpha \sim 0.2$ obtained 
from the complete sample. The actual value is consistent
with this, giving another confirmation 
that our model contains all the main ingredients needed to explain 
g-g lensing. All the reduced $\chi^2$ values are around 1, so the model fully 
explains the data.

We can test the assumption that for a given luminosity the
halo mass and its profile does not depend on the enviroment 
by comparing the $M_{\star}$ between high and low density samples.  
If the galaxies inside larger halos do not keep any dark matter 
one should see a reduction in the high 
density sample at small radii, since for $1-\alpha=0.6$ of all the
galaxies there would be no individual halo attached to them. 
Such a model would give fitted $M_{\star}$ in the high density sample 
about 60\% of that in the low density sample.
In fact
we find that the best fitted value for $M_{\star}$ are comparable,
which argues against this assumption. While the errors are still large
and the 
current data does now allow us to test this hypothesis in more detail
we conclude that there is no evidence for a significant halo stripping
or different star formation efficiency
for galaxies within larger halos. This is consistent with the expectations 
from the 
noninteracting cold dark matter models \cite{2000ApJ...544..616G}, but may not be with 
self-interacting dark matter models which predict significant 
mass loss in subhalos \cite{2000PhRvL..84.3760S}. 

\begin{figure}
\begin{center}
\leavevmode
\epsfxsize=10.0cm \epsfbox{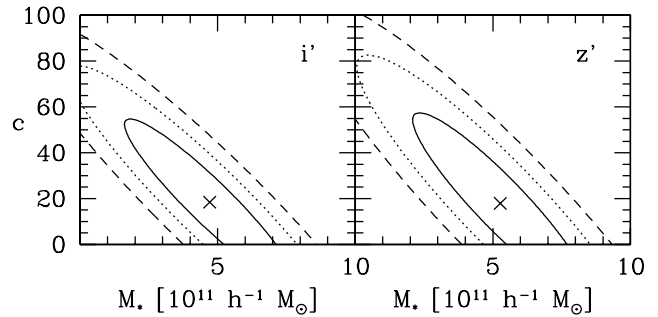}
\end{center}
\caption{Contour plots in $c-M_{\star}$ plane for $i'$ and $z'$. While the 
two parameters cannot be determined separately there is a weak 
dependence of $M_{\star}$ on $c$.
}
\label{fig17}
\end{figure}

Since we have a signal as a function of radius we should be able 
in principle to place limits on the concentration 
parameter $c$ independent of the halo mass. 
In practice broad luminosity distribution and significant clustering 
correction prevents us from doing so. We have tried to do a formal 
fit to $M_{\star}$ and $c$ simultaneously and the resulting contour plots 
show a strong degeneracy between the two parameters  
(figure \ref{fig17}). 
Given that we cannot determine $c$ independently we must asses its 
influence on $M_{\star}$. As seen in figure \ref{fig17} 
there is some weak anti-correlation of $M_{\star}$ 
with $c$. This arises because higher $c$ increases the signal in the 
first bin, which has a significant contribution to the overall fit.
This correlation can be parametrized as $M_{\star} \propto c^{-0.15}$ and since 
$c$ is expected to be in the range between 8 and 15 this uncertainty 
affects $M_{\star}$ by 10\% at most and so does not dominate the error budget. 

\subsection{Morphology dependence}

\begin{figure}
\begin{center}
\leavevmode
\epsfxsize=10.0cm \epsfbox{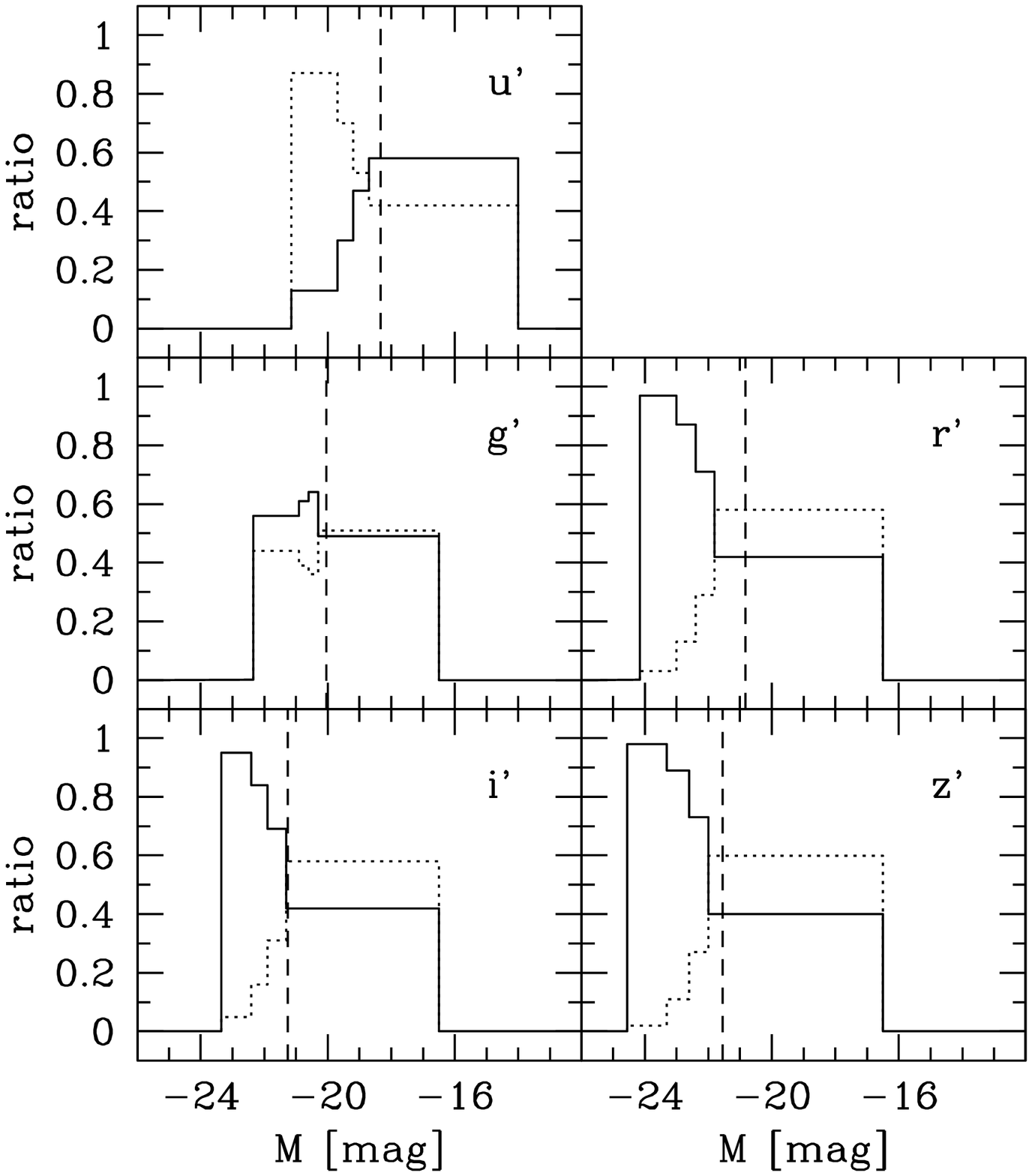}
\end{center}
\caption{Fraction of early types (solid) and late types (dotted) in 
each of 4 luminosity bins for all 5 colors. Also shown is the magnitude of 
$L_{\star}$ galaxy (dashed).
}
\label{fig14}
\end{figure}

So far we have assumed that morphology of the lens galaxy does 
not affect the relation between 
luminosity and mass, so that we 
do not have to worry about their differences in a sample which 
mixes early and late types. This is a reasonable assumption 
in the red/infrared bands where light is supposed to be a reliable
measure of the total stellar mass,
assuming the morphology of a galaxy (which is believed to be related to the 
merging history) is 
not correlated with the amount of gas in a halo that cools to form stars. 
Observationally this was suggested by the 
SDSS team \cite{2001astro.ph..8013M}, showing 
that mass to light ratios at a fixed radius in the reddest
bands were very similar.
This analysis
did not account for the contribution from groups and clusters and for
observed mass-luminosity scaling.
Here we assume mass luminosity relation based on $\beta$ values in 
table \ref{table1}.
We use the sample split into  
late and early types that
has about equal number of galaxies for the two types. 
The morphological  
classification is based on the color information and light concentration 
index and the early type sample is dominated by ellipticals, but 
also includes S0 and  
Sa galaxies, while the late type sample is dominated by Sb/Sc's
(see \shortciteNP{2001astro.ph..7201S} for a thorough 
discussion of the relation to the Hubble sequence). 
Figure \ref{fig14} shows the fraction of the two types as a function of 
luminosity for the 4 bins used here. It is clear that at the 
bright end in $r'$, $i'$ and $z'$, 
the sample is dominated by the early types, while in $u'$ it is just the 
opposite. In $g'$ the fraction is roughly independent of luminosity.
Note that the bright end is where most of the g-g lensing signal is.

\begin{figure*}
\begin{center}
\leavevmode
\epsfxsize=10.0cm \epsfbox{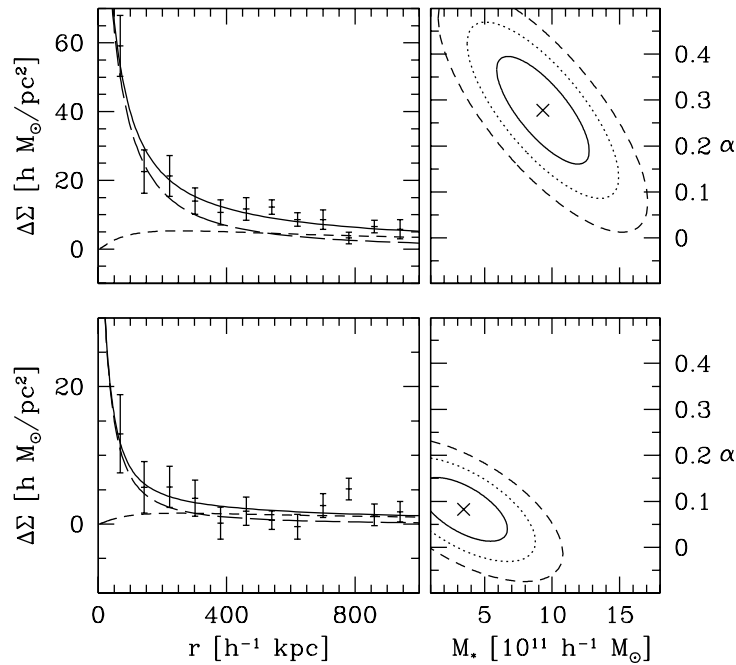}
\end{center}
\caption{Top: best fitted model to the early type sample together 
with the contour plot in $M_{\star}-\alpha$ plane for luminosity 
distribution in $i'$. 
Bottom: same for the late type sample. 
}
\label{fig15}
\end{figure*}

We fit for 
$M_{\star}$ and $\alpha$ in the two subsamples. The results are given in 
table \ref{table2} for all 5 colors.
We do not apply any internal extinction correction 
which should be significant for
the late type sample in the blue bands, 
so for those the intrinsic luminosities could be even 
brighter. In addition, while Petrosian magnitudes used by the SDSS 
should account for most of the light from the late types with exponential 
profiles they may underestimate
the light from the early types with de Vaucouleurs profile by up to 20\%.
Figure \ref{fig15} shows the best 
fitted model together with the data. We find $\alpha \sim 0.07 \pm 0.04$ 
for the late type sample,
showing that most of the late type galaxies are in the field. 
For the 
early type galaxies we find $\alpha \sim 0.28 \pm 0.06$, 
so a significantly larger 
fraction of these is in groups/clusters. This is another evidence of 
morphology-density relation, although in this case we are not probing 
the relation as a function galaxy density \cite{1980ApJ...236..351D},
but rather the fraction of
galaxies around $L_{\star}$  
that are in groups of a broad range in mass around $10^{13}M_{\sun}$.
Groups of such low mass are very difficult to study observationally
and our approach is one of a few ways to study the galaxy 
membership in these halos.
The errors quoted above are only statistical and the parameter 
$\alpha$ depends also on the assumed values for 
$\epsilon$, $M_{\rm cutoff}$ and $y_g$ as discussed in previous 
section. We estimate the systematic error on $\alpha$ 
should be of order 30\%.

The values for $M_{\star}$ differ significantly between early and late types
in all bands. The difference is almost a factor of 10 in $u'$, 
6 in $g'$, dropping 
to a factor of 2.5 in $r'$ and 2 in $i'$ and $z'$. 
This differs somewhat from the conclusions 
in \shortciteN{2001astro.ph..8013M}, where it was found that $M/L$ for the two types 
becomes equal in $z'$. The main reason for the difference is that we 
assume $M \propto L^{\beta}$ with $\beta$ values from table \ref{table1}
and that we apply 
luminosity dependent clustering correction to the data, which boosts 
the brighter early type sample more than the fainter late type sample. 
However, the error on $M_{\star}$ for the
late types is large, since the signal is rather weak, and in 
$i'$ and $z'$ the data are still consistent with the assumption of 
equal $M/L$ ratios for the two types. In addition, Petrosian 
magnitudes may miss up to 20\% of light for early type profiles and much 
less than that for late types, so this may bring the mass to light 
ratios closer by this amount. 
Theoretically one would 
expect $M/L$ to be quite similar if $i'$ and $z'$ luminosity is a reliable 
tracer of stellar mass and if star formation efficiency is independent 
of morphology. The color comparisons indicate that the
difference in luminosity between early and late types is only 20-30\%
from $i'$ to K band (which should be closest to tracing stellar mass).
We discuss the validity of these assumptions further below. 

Given that the $M/L$ differs significantly between early and late 
types and that their relative contribution changes with 
luminosity we must revisit the mass-luminosity scaling of 
previous subsection. One possibility is to use $g'$, where the two 
populations are in equal ratios across all luminosity bins. 
Since at a given luminosity late types are much less massive 
the dominant contribution comes from the early types, so the deduced 
$\beta=1.3-1.7$ is the correct scaling for early types regardless
of $\beta$ for late types.
In $i'$ and $z'$ the early types dominate at the brightest end
and contribute only 40\% to the faintest bin, so here a 
morphological dependence of
$M/L$ could induce a change in $\beta$. On the other hand, in these bands
the late types are only a factor of 2 less massive for a 
given luminosity. This gives 
$\beta$ for early types about 0.15 lower than the value 
found for the 
overall population. Similar values are found also in $r'$ and $z'$. 
We conclude that $\beta$ for early types is in the range $1.4 \pm 0.2$ in 
the red bands.

Variation of morphology with luminosity
has a larger effect on the correlation between luminosity and 
mass in $u'$. Here late types dominate at the bright end and early types
at the faint end, so the massive galaxies are actually on average fainter than 
the less massive ones. If we use $\beta=1.4$ for early types together with 
the
scaling of early type fraction with luminosity (figure \ref{fig14})
we obtain  
$\beta \sim 0.5$, in agreement with the fitted value in this band.
Lack of correlation in this band is therefore not necessarily a 
consequence of $u'$ light being uncorrelated with mass, but rather 
of a changing morphological fraction with luminosity coupled
to a large difference in $M/L$ at a given $L$ in that band.

The current data sample does not allow for a determination of
mass luminosity relation for late types. 
We have assumed the same value of $\beta$ for both types here, but it 
could well be that $\beta$ is lower for the less luminous and less 
massive late type sample, as suggested 
by the semi-analytic models of galaxy formation (\shortciteNP{1999MNRAS.303..188K}, \shortciteNP{2000MNRAS.311..793B}). If we use 
$\beta=1$ for the late type the best fitted $M_{\star}$ does not change 
much, so the difference in $M/L$ between early and late type sample 
cannot be explained by this effect. 

\begin{figure}
\begin{center}
\leavevmode
\epsfxsize=10.0cm \epsfbox{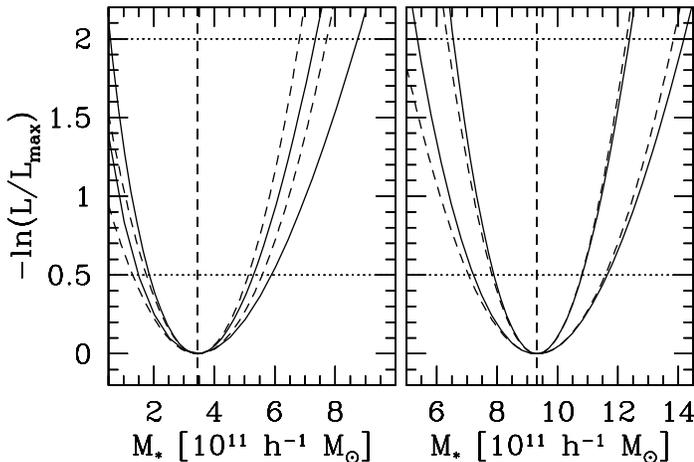}
\end{center}
\caption{Comparison between exact log-likelihood (solid) and its gaussian 
approximation assuming rms values from the nonlinear fitting routine 
(dashed). Left is for late type and right for early type galaxies. 
Narrower curves are for a fixed value of $\alpha$ at the minimum, 
while broader
curves are marginalized over $\alpha$. In both cases, gaussian 
distribution gives a good description of the error probability 
distribution.  }
\label{fig22}
\end{figure}

The errors in tables \ref{table1}, \ref{table2} are assumed to be 
gaussian in $M_{\star}$, $\alpha$ and $\beta$. For $M_{\star}$ we show the comparison
in figure \ref{fig22}, where the log-likelihood 
is plotted against the gaussian approximation for early and late
type galaxies. We show both the case 
assuming a constant $\alpha$ (where the rms error is given by the 
inverse of corresponding diagonal term in the curvature matrix) and 
marginalized over $\alpha$ (where the rms error is given by the 
corresponding diagonal term in the inverse of curvature matrix). 
In all cases
the gaussian approximation gives a good description
of the true likelihood at least out to 2-sigma level. This is a consequence of 
the fact that the signal scales almost linearly with the virial mass $M_{\star}$
(figure \ref{fig4}).

Since for early types the group and cluster contribution is significant 
the values of $M_{\star}$ depend somewhat on the assumed radial 
distribution of galaxies and dark matter, as well as on the halo 
occupation statistics. We find that if the group and cluster profile
is shallower then fitted values do not change much, while if they are
more concentrated then part of the signal at smaller radii can be 
attributed to group/cluster contribution and this lowers $M_{\star}$. For 
example, assuming $c=10$ both for group/cluster dark matter and
for radial galaxy distribution results in a 25\% reduction of $M_{\star}$ 
for early types, while the corresponding value for late types changes
significantly less. This is probably an upper limit, since $c=10$ is 
a more appropriate value for galaxy sized halos rather than groups and 
clusters, but one should nevertheless keep in mind this as an additional 
source of systematic error on $M_{\star}$.

\section{Discussion}

This paper establishes a quantitative framework on how to analyze g-g 
lensing data and applies it to the SDSS sample of 35,000 galaxies 
with known redshifts 
and luminosities. The main qualitative new 
feature of the model is that 
both the dark matter around the individual 
galaxies and dark matter in groups and 
clusters are included. Both components are needed
to explain the observations. 
The two contributions have a different radial dependence and 
can be determined separately. This provides important constraints 
on the galaxy formation models, which must satisfy the relative
contribution from central and group/cluster components.
We also argue that
correlations between the  
galaxy and another halo that is not the one that galaxy belongs to 
can be neglected, at least on scales below 1$h^{-1}$Mpc of interest here.

The main
result of this paper is determination of 
galaxy virial masses and the fraction
of galaxies in groups and clusters as a function 
of luminosity and morphology.
These provide important 
constraints on the galaxy formation models. 
The average
virial mass $M_{200}$ of an $L_{\star}$ galaxy  
is around 
$(5-10)\times 10^{11}h^{-1}M_{\sun}$, depending on the passband. 
This mass varies significantly with morphology and the variation is 
largest in $u'$, where the difference between early and late 
types can be up to a factor of 10,
decreasing to a factor of 2-3 in $r'$, $i'$ and $z'$. 
For example, in $i'$ we find $M_{200}=(3.4 \pm 2.1) \times 10^{11}h^{-1}M_{\sun}$
for late types and $M_{200}=(9.3\pm 2.2) \times 10^{11}h^{-1}M_{\sun}$
for early types.
While the signal for late types is rather weak and consequently the errors are 
large, they are gaussian distributed and so
we can exclude the possibility that in red bands the $M_{\star}$ 
for early and late types are equal.

What does this imply for the star formation efficiency as
a function of galaxy morphology? To address this we first transform from 
$i'$ to K band, where the luminosity is assumed to be 
a reliable tracer of stellar mass for a given age of the population (and IMF). 
This transformation does not change the results significantly, since 
$K-i'$ differs by 0.2-0.3 magnitudes at most 
between the two types \shortcite{2001astro.ph.11024I}.
This difference is further reduced by up to 20\% because of the
missed light for the de-Vaucouleurs profile. 
As a result, the difference in $M/L$ between the two types 
is reduced to a factor of 2. In K band the difference in $M_{\rm stellar}/L$
between early and late type population can be up to a factor of 2
depending on the exact ages and IMF assumed, so our results are
consistent with the assumption that
star formation efficiency for early and late type galaxies is the same. 
The errors however are still large both on the virial masses 
and stellar mass to light ratios and 
there could still be up to a factor of 2 difference. 
Since the current results are based on only 5\% of final SDSS sample 
one should be able to place much better constraints on this issue in 
the future.

If we adopt the late(early) type stellar mass to light ratio $M_{\rm stellar}/L
\sim 1.5(3)h$ in $i'$ 
we find that at $L_{\star}$ about 10-15\% of 
virial mass is converted to stars.
The fraction of baryons inside a halo 
should equal $\Omega_b/\Omega_m$, which can vary between 
10-20\% for $\Omega_m \sim 0.3 \pm 0.1$ and $\Omega_b \sim 0.04 \pm 0.01$.
We see that if our halo masses are correct a significant
fraction of baryons, up to 100\%, is converted to stars in such halos. 
For early types this fraction decreases with luminosity and 
is a factor of 2 lower 
at $7L_{\star}$.
These stellar fractions 
can be accomodated in the standard models, but 
require a low matter density and/or a high baryon density, so that the 
overall baryon to dark matter fraction is sufficiently high.

We find that $M/L \propto L^{0.4\pm 0.2}$ in red bands ($i'$ and $z'$).
An increase in $M/L$ with luminosity above $L_{\star}$ is 
expected theoretically from semi-analytic models of galaxy formation
(\shortciteNP{1999MNRAS.303..188K}, \shortciteNP{2000MNRAS.311..793B}).
Comparison of these results with mass tracers such as 
Tully-Fisher or fundamental plane is complicated, because these  
probe the halo at smaller radii, where the rotation velocity of dark matter 
could be different and where baryons could play a significant role. 
A detailed analysis will be presented elsewhere. The results from 
such analysis show
that the masses obtained here are consistent with the CDM picture 
in which the rotation velocity 
drops by roughly 1.7-1.8 from the optical to 
the virial radius at $L_{\star}$.
Such a drop is expected in CDM models, where the 
rotation curve peaks at a fraction of the virial radius and 
drops beyond that. However, to explain such a large drop one also needs 
a significant additional contribution to the rotation velocity 
from the stellar component and/or baryon compressed dark matter.
For early types the scaling $M\propto L^{1.5}$
indicates $L \propto v_{200}^{2}$, while for the same galaxies 
the Faber-Jackson relation 
gives $L \propto \sigma^{4}$ \cite{2001astro.ph.10344B}. In this case the ratio of optical to 
virial circular velocity depends on luminosity
and drops to 1.4 at $7L_{\star}$.

Comparison with other g-g lensing results is also complicated, since 
none of these use luminosity information, are typically at a higher 
redshift with poorly determined distances 
and may have a different morphological composition. The 
most direct comparison can be made to 
the study of early type galaxies by \shortciteN{2001ApJ...555..572W}, where 
color information was used to determine approximate redshifts to these 
galaxies. 
A direct comparison shows that their virial mass for early type 
$L_{\star}$ galaxy is around $2\times 10^{12}h^{-1}M_{\sun}$.
Their sample is at a higher redshift and may not be directly 
comparable to ours, but the obtained value is quite close to our best fitted 
value of
$M_{\star}=2.1\times 10^{12}h^{-1}M_{\sun}$ in $g'$, which is
the closest to $B$ band used there (note that $L_{\star}=1.1\times 10^{10}h^{-2}
L_{\sun}$ in both samples). In their analysis they
assume $\beta=0.5$, which is 
inconsistent with the
SDSS luminosity dependent data, so their actual value of $M_{\star}$ could even 
be somewhat lower because $\beta$ and $M_{\star}$ are anticorrelated.
Earlier analysis of the dynamics of satellites around spiral galaxies 
gave significantly higher masses (up to 
a factor of 5, \shortciteNP{1997ApJ...478...39Z}) 
and also did not show correlation 
between light and mass, but a more recent analysis
of the SDSS data shows a better agreement with g-g lensing results 
(T. McKay, private communication). 
At the upper end of the luminosity range our results can be compared to 
group and cluster velocity dispersion analysis of 
\shortciteN{2001astro.ph.12534G}. Our results agree well
both on the mean luminosity 
of a $10^{13}h^{-1}M_{\sun}$ halo and on the scaling of mass with luminosity.
For late type galaxies at $L_{\star}$ our results agree well with the virial 
mass derived from theoretical models by \citeN{2001astro.ph.12566V}.

There are many aspects of the analysis that could be improved upon with 
better statistics which will be available in the future. 
We mention some here.
We have assumed a power law
relation between mass and luminosity with a constant slope $\beta$, while 
theoretical models suggest $\beta$ changes with mass. 
With the current sample
the signal for galaxies less luminous than $L_{\star}$ is too small
to be detectable and the relation between mass and luminosity cannot 
be established in that range. As larger samples become available this 
regime will be probed as well.
Fraction of galaxies in groups and clusters should be determined as 
a function of luminosity.
Similarly, galaxy morphology dependent analysis should be improved by 
analyzing luminosity dependence of the signal for each morphology type
and by dividing the lens galaxies into several morphological classes.
With better statistics 
the fraction of galaxies in groups and clusters could be 
determined as a function of luminosity and not just for the overall 
sample as done here. 
The 
radial distributions of dark matter and galaxies in groups and clusters 
as well as galaxy occupation statistics as a function of halo mass
should be quantified better. 
These effects
can change the values for $M_{\star}$ and $\beta$ by 20-30\% and remain 
the dominant source of systematic error. 
Better k-correction 
and evolution effects should be applied to the data to match
the bright end (which receives contributions from  $z \sim 0.2$) 
and faint end (low $z$) samples. This effect should not exceed 0.1-0.2 
magnitudes in red bands and so should not change our conclusion on the mass
luminosity relation significantly.
Differences between halos of
galaxies in the field and in groups and clusters could be studied
in more detail, testing for example the collisionless CDM paradigm.
In this paper we assumed that the profiles of galaxies within groups 
and halos are unchanged out to the truncation radius. In the extreme 
opposite case where galaxies inside groups and clusters do not retain 
any mass at all the virial masses of galaxies in the field will be 
underestimated by $1-\alpha$. This is a small correction for late type 
galaxies, but potentially more important for early type galaxies. 
Finally, the clustering correction, which corrects for the fraction 
of galaxies that are correlated with the lens and are not in the background,
is luminosity dependent and should be done more accurately. 
While the 
procedure adopted here is reasonable, it should 
be studied in more detail. This
effect may be removed for example if one can select background galaxies
using their photometric redshift (photo z) information, although it
is not clear if the faint galaxies that dominate the background population
have sufficient signal to noise for this purpose. 
Current experience with photo z's indicates that in
SDSS these work only down to $r'<20.5$, while most of the background 
galaxies used for shape information are fainter than that. 

As we obtain more data better statistical analysis will also be needed.
Rather than divide the data in arbitrary 
luminosity bins, which may still be too broad for a quantitative analysis,
one should parametrize the model with a few parameters and 
use maximum likelihood type of analysis to determine these.
Photometric redshifts would 
also help to improve the statistics since one could weight the signal 
by giving optimal weight for each lens-background galaxy pair (this would 
then automatically downweight all the galaxies that are close to 
the lens). These improvements will allow one to make very robust 
statements on the relation between the virial mass and luminosity 
and on the
membership in groups and clusters 
over a broad range of luminosities and morphological types. 

US acknowledges he support of NASA, David and Lucille
Packard Foundation and Alfred P. Sloan Foundation. 
J.G. was supported by grants 5P03D01820 and 2P03D01417 from Polish State
Committee for Scientific Research.
We thank 
Tim McKay, Erin Sheldon and 
Iskra Strateva 
for help with the interpretation 
of SDSS data and its analysis.

\bibliography{apjmnemonic,cosmo,cosmo_preprints}
\bibliographystyle{mnras}

\onecolumn

\begin{table}
		\begin{center}
		\begin{tabular}[t]{c r @{$\pm$} l r @{$\pm$} l r @{$\pm$} l c r @{$\pm$} l} 
			\hline
			\hline
			band & \multicolumn{2}{c}{$M_{\star} [10^{11} h^{-1} M_{\odot}]$ } & 
        		\multicolumn{2}{c}{$\beta$} & \multicolumn{2}{c}{$\alpha$} &
				\multicolumn{1}{c}{$R_{M_{\star}\beta}$} &
 				\multicolumn{2}{c}{$L_{\star} [10^{10} h^{-2} L_{\odot}]$} \\
			\hline 
			\hline
 $u'$ &   8.05 &  1.23 &  0.23 &  0.22 &  0.23 &  0.03 & -0.04 &  0.78 &  0.06 \\
				\hline
 $g'$ &  10.24 &  1.62 &  1.67 &  0.22 &  0.20 &  0.03 & -0.71 &  1.11 &  0.04 \\
				\hline
 $r'$ &   8.96 &  1.59 &  1.51 &  0.16 &  0.17 &  0.03 & -0.83 &  1.51 &  0.04 \\
				\hline
 $i'$ &   8.54 &  1.53 &  1.51 &  0.14 &  0.17 &  0.03 & -0.85 &  2.05 &  0.08 \\
				\hline
 $z'$ &  10.01 &  1.66 &  1.44 &  0.13 &  0.16 &  0.03 & -0.83 &  2.58 &  0.10 \\
				\hline
				\hline
 $u'$ &   6.16 &  1.70 & -0.52 &  0.29 &  0.15 &  0.02 &  0.83 &  0.78 &  0.06 \\
				\hline
 $g'$ &   9.33 &  1.58 &  1.16 &  0.23 &  0.17 &  0.02 & -0.35 &  1.11 &  0.04 \\
				\hline
 $r'$ &   7.56 &  1.55 &  1.34 &  0.17 &  0.18 &  0.03 & -0.72 &  1.51 &  0.04 \\
				\hline
 $i'$ &   7.16 &  1.53 &  1.40 &  0.15 &  0.17 &  0.03 & -0.79 &  2.05 &  0.08 \\
				\hline
 $z'$ &   8.08 &  1.62 &  1.35 &  0.14 &  0.17 &  0.03 & -0.75 &  2.58 &  0.10 \\
				\hline
				\hline
			\end{tabular}

 \caption{Best fitted parameters to the SDSS data in five colors 
assuming $\alpha=0$ for all but the faintest bin. The corresponding values 
for $M_{\star}$ using equal $\alpha$ for all bins are 10-20\% lower, while 
the other parameters do not change much. Also 
given is the correlation 
        coefficient between $M_{*}$ and $\beta$ and the value of $L_{\star}$.
\label{table1}}
		\end{center}
\end{table}

\begin{table}
        \begin{center}
                \begin{tabular}[t]{c c r @{$\pm$} l r @{$\pm$} l r @{$\pm$} l }
                        \hline
                        \hline
                        band & type &\multicolumn{2}{c}{$M_{\star}$ [$10^{11} h^{-1} M_{\odot}$]} 
                        & \multicolumn{2}{c}{$\alpha$} & \multicolumn{2}{c}{$L_{\star}  [10^{10} h^{-2} L_{\odot}]  $} \\
                        \hline
                        \hline
 $u'$ & E  & 25.55 &  5.78 &  0.31 &  0.07 &  0.78 &  0.06 \\

			\hline
 $u'$ & S  &  2.52 &  1.58 &  0.09 &  0.04 &  0.78 &  0.06 \\

			\hline
 $g'$ & E & 21.94 &  5.25 &  0.29 &  0.07 &  1.11 &  0.04 \\
			\hline
 $g'$ & S &  3.18 &  2.02 &  0.08 &  0.05 &  1.11 &  0.04 \\
			\hline
 $r'$ & E & 10.73 &  2.53 &  0.28 &  0.08 &  1.51 &  0.04 \\
			\hline
 $r'$ & S &  3.26 &  2.08 &  0.08 &  0.05 &  1.51 &  0.04 \\
			\hline
 $i'$ & E &  9.32 &  2.26 &  0.28 &  0.08 &  2.05 &  0.08 \\
			\hline
 $i'$ & S &  3.44 &  2.14 &  0.08 &  0.05 &  2.05 &  0.08 \\
			\hline
 $z'$ & E  & 10.06 &  2.38 &  0.28 &  0.08 &  2.58 &  0.10 \\
			\hline
 $z'$ & S &  4.08 &  2.63 &  0.08 &  0.05 &  2.58 &  0.10 \\
			\hline
                        \hline
                \end{tabular}
        \caption{Best fitted parameters 
for early type (E) and late type (S) galaxies 
as a function of color. Early types have higher $M_{\star}$ and group/cluster 
fraction than late types. 
\label{table2}
}
        \end{center}
\end{table}

\twocolumn
\end{document}